\pdfoutput=1
\documentclass[a4paper,withtimes]{easychair}
\usepackage{fullpage}
\usepackage{wrapfig}

\newcommand{\coqpsv}{coq-psv}

\title{The Coq Proof Script Visualiser (coq-psv)%
\thanks{Special thanks to \href{http://www.cs.uni-potsdam.de/\~sboehne}{Sebastian Böhne} who gave the idea and concept to the tool.}~~\footnote{This work was presented during a talk at the \href{https://coq-workshop.gitlab.io/2020}{Coq Workshop 2020}}}

\author{
Mario Frank
}

\institute{
  University of Potsdam,
  Institute of Computer Science,
  Potsdam, Germany\\
  \email{mario.frank@uni-potsdam.de}, \href{https://orcid.org/0000-0001-8888-7475}{orcid:0000-0001-8888-7475}  
 }

\authorrunning{Frank, M.}

\titlerunning{The Coq Proof Script Visualiser}

\begin{document}

\maketitle



\setcounter{tocdepth}{2}

%
%




The Coq proof assistant~\cite{coq} is used in various scientific, industrial, and teaching contexts and enables scientists, teachers, and students to exchange formalisations in
an easy way.
But this exchange is limited to the vernacular files and the respective \LaTeX{} representation generated by coqdoc~\cite{coqdoc}, for example. Also, those representations merely contain the definitions and proof scripts but not the information about the specific proof steps, i.e.\,the current goals and hypotheses. Thus, in order to fully understand the specific proofs, including the way how hypotheses change, the recipient needs to acquire a compatible version of Coq.
This can be cumbersome as the installation of the specific Coq version and
necessary additional tools is not always straight-forward due to operating system
and Opam version differences. To our knowledge, there currently is no tool that is able to represent
Coq proofs on paper including all used tactics, goals and hypotheses in a readable and printable manner as tree representations usually do not scale well for long proofs.


\paragraph{Applications.}

The Coq Proof Script Visualiser aims at supporting the development, teaching, and review process of proofs by exporting the proof scripts together with the current proof state for each proof step. As all information about the proof is then at hand, readers merely need to know the semantics of the used tactics to understand the proof. Proof developers can benefit from additional warnings given by \coqpsv{} when superfluous structuring is used. Reviewers can retrace the proof more easily without the need to install Coq. In teaching, the generated output can be easily modified to be used as a cloze where students have to fill in the matching tactic or introduced hypothesis, for example.
And most notably, the exported proof representations can be used for archiving purposes and enriching the output of coqdoc.


\paragraph{Approach.}

Given a Coq file, \coqpsv{} processes it using the Coq parsing functionality and data structures. Although the output of \coqpsv{} is linear, it internally generates a labelled proof tree for each proof in the file as using a tree structure simplifies the analysis and augmenting of the proof information. The label itself contains the name and type of the proven statement (e.g.\,theorem or lemma) together with the statement itself and the information, how the proof was completed (Qed or Admitted). The tree consists of nodes that contain the used tactic or command, and the resulting hypotheses and goals. While the extraction of the label content is easy, building the tree nodes is much more complex. This is due to the numerous ways in Coq to branch in a proof situation, which includes bullets, brackets and the (deprecated) Focus command.
After extracting the labelled trees, they are translated to \LaTeX{} tables according to the given command line options. We will give insight into the way we handle different branching options and why we do so in the talk.


\begin{wrapfigure}{r}{0.5\textwidth}
  \vspace{-2.35em}
  \begin{center}
    \includegraphics[width=0.49\textwidth]{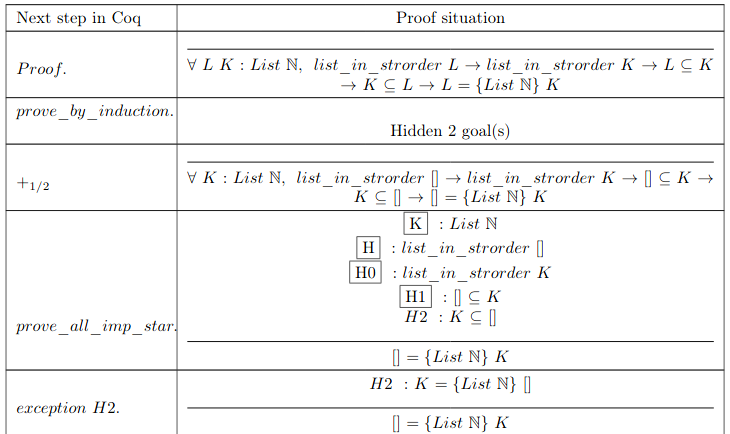}
  \end{center}
  \caption{Excerpt of a proof visualisation}
  \label{img:screenshot-coq-proof}
\end{wrapfigure}
\paragraph{Versatile Use.}

The coq-psv supports many command line options that define the output behaviour. It is able to process both single Coq files and complete Coq projects. Furthermore, it is possible to generate either complete \LaTeX{} documents or only the \LaTeX{} tables. Also, all proofs from a vernacular file can be exported either to one single file or to separate files per proof. The latter simplifies the integration into other \LaTeX{} documents.
There are two different table layouts: The first one mimics the goal and hypothesis visualisation of CoqIDE while the second one represents the goals and hypotheses in sequent style. Also, it is possible to hide goals and hypotheses in the step in which they are created if all of them are handled by bullets.
This can reduce the size of the output significantly.
Hiding hypotheses that remain unchanged until the end of the (sub-)proof and marking them as invariant on introduction is also supported, which again condenses the size of the tables (see Figure \ref{img:screenshot-coq-proof}).
Finally, the layout of the tables can be modified after the generation. This can be done by overriding the default \LaTeX{} commands used in the generated files before importing them. For example the size of the table itself and of specific table columns can be set to a desired value without having to modify the table itself.


\paragraph{Related work.}

There are already some approaches to representing proof scripts in a more readable and independent way (see~\cite{hendriks2010,kawabata2018,libal2014,pais2013,proviola2010,ProofTree}, for example). But except for~\cite{hendriks2010} and \cite{proviola2010}, all of them represent the proofs only as trees or diagrams. While trees and diagrams can give a good overview about the structure of a proof, printing them in a readable way can be problematic for long proofs. Although the approach described in~\cite{hendriks2010} supports both a tree and a Fitch style representation, the latter only shows the current step and it does not seem to be possible to export the complete proof as \LaTeX{} file, for example. Also, only ProofWeb~\cite{hendriks2010}, Proviola~\cite{proviola2010}, ProofTree~\cite{ProofTree} and Traf~\cite{kawabata2018} are targeting Coq and the two latter only work in a live session with Coq and, for example, ProofGeneral~\cite{ProofGeneral}.
Furthermore, Traf uses ProofTree and both seem to have been discontinued for more than two years. 
according to the project pages~\cite{ProofTree,Traf}. Proviola is able to generate an animated version of a proof script giving good insight into the proving process, but it does not focus on printing the output on paper. Anther project makes Coq proofs available in the browser (see~\cite{www:jscoq}) which makes the local installation of Coq unnecessary and improves usability as documents can be shared. But it does not seem to be possible to generate a printable version with all necessary information as coq-psv does.

\paragraph{Current state and future work.}

The tool should be currently seen as proof of concept and supports Coq 8.10. It is implemented in OCaml and can be installed from our Opam repository as described on the \href{http://www.cs.uni-potsdam.de/coq-psv}{\coqpsv{} project page}~\cite{www:coq-psv}. There are currently some minor issues in the vertical alignment of the used tactic. But even non-trivial proofs can be visualised decently as can be seen in some examples on the project page. It is possible to generate \LaTeX{} and PDF files. Obviously, there are many options for future work. For example, one could hide tactics like clear, rename and move and give them a special treatment instead: for instance, marking cleared hypotheses, annotating identifier changes with the substitution pattern 
and just change the position of the displayed hypotheses. Then, a support for modules in files should be incorporated. Furthermore, supporting especially HTML output is a main aspect of future work. We aim at integrating such an improved \coqpsv{} into coqdoc.
Additionally, \coqpsv{} could be extended to improve the proof scripts themselves. For instance, structure improvements can be suggested. Especially, unused or irrelevant hypotheses might be detected and automatically deleted, if desired. This is not an easy task since it may change the names of the hypotheses appearing later on. Finally, one could give the option to conduct an automatic reordering of the lines in a proof script.

\bibliographystyle{plain}
\bibliographystyle{abbrv}
\bibliography{coq-psv}

\end{document}